\documentclass[prb,amsmath,amssymb,superscriptaddress,twocolumn]{revtex4}
\usepackage{float}
\usepackage{amsmath}
\usepackage{amssymb}
\usepackage{amsfonts}
\usepackage{euscript}
\usepackage{enumerate}
\usepackage{hhline}
\usepackage{pslatex}
\usepackage{tabularx}
\usepackage[usenames,dvipsnames]{xcolor}
\usepackage{graphicx}

\usepackage{dcolumn}
\usepackage{bm}
\usepackage[sort&compress]{natbib}

\usepackage{lipsum}

\newcommand{\blfootnote}[1]{%
  \bgroup
  \renewcommand{\thefootnote}{\fnsymbol{footnote}}
  \footnotetext[0]{#1}
  \egroup
}

\makeatletter
\renewcommand{\@biblabel}[1]{#1. }
\renewcommand{\@dotsep}{500}
\renewcommand{\@pnumwidth}{0em}
\renewcommand{\l@figure}[2]{
\@dottedtocline{1}{1.5em}{2em}{Figure #1}{}\vspace{15pt}}

\usepackage[normalem]{ulem}

\begin{document}
\title{Photogalvanic second harmonic generation in Si$_3$N$_4$ for 1 Hz level on-chip metrology and spectroscopy} 




\author{Andrei Diakonov}\email{andrei.diakonov@mail.huji.ac.il}
\affiliation{Institute of Applied Physics, The Hebrew University of Jerusalem, Israel}
\author{Roy Zektzer}
\affiliation{Alexander Kofkin Faculty of Engineering, Bar-Ilan University, Israel}
\author{Xiyuan Lu}
\affiliation{Microsystems and Nanotechnology Division, Physical Measurement Laboratory, National Institute of Standards and Technology, Gaithersburg, MD 20899, USA}
\affiliation{Joint Quantum Institute, University of Maryland College Park, College Park, MD 20742, USA}
\author{Kartik Srinivasan}
\affiliation{Microsystems and Nanotechnology Division, Physical Measurement Laboratory, National Institute of Standards and Technology, Gaithersburg, MD 20899, USA}
\affiliation{Joint Quantum Institute, University of Maryland College Park, College Park, MD 20742, USA}
\author{Liron Stern}
\affiliation{Institute of Applied Physics, The Hebrew University of Jerusalem, Israel}

\date{\today}

\begin{abstract}

    The coherent photogalvanic (PG) effect induces an effective $\chi^{(2)}$ nonlinearity in the widely available but natively $\chi^{(3)}$ silicon nitride integrated photonic platform, unlocking a pathway toward chip-scale precision spectroscopy and optical clockworks. In the context of second harmonic generation (SHG), the underlying physics relies on the induction of an internal electric field whose character depends on the phase matching approach. When the fundamental and second harmonic waves are directly phase-matched, typically through an intermodal scheme, a spatially uniform electric field is created. When there is a phase-mismatch between fundamental and second harmonic, a spatially-varying electric field is generated, and enables quasi phase-matching. Even though the quasi phase-matching approach is highly flexible and tunable, it has been shown that it can be accompanied by an offset of the second harmonic frequency, dependent on the pump power and detuning. This raises the question of whether direct phase matching can support metrologically compatible SHG. Here, we test the preservation of the (2{:}1) frequency ratio in directly phase-matched PG-SHG by comparing the fundamental and doubled optical frequencies in a silicon nitride microresonator. We find a $<1$~Hz frequency offset, in contrast to previous observations in quasi-phase-matched PG-SHG. Furthermore, we measure a residual fractional frequency instability of the PG-SHG induced noise of $2\times10^{-15}$ at 1~s, averaging down to the $10^{-16}$ level at 1000~s, with deviations from exact doubling below 1 Hz over multi-hour operation, serving as an upper bound on any additional instability introduced by the PG-SHG process. Together, these results establish directly phase-matched PG-SHG as a metrologically compatible route to effective $\chi^{(2)}$ functionality in silicon nitride, combining sub-Hz frequency-ratio fidelity with low residual instability and high coherence on a mature integrated photonics platform for optical clockwork, self-referencing, and precision spectroscopy.
\end{abstract}

\maketitle

\blfootnote{$^{\ddagger}$~This document is preliminary and is intended for peer review conducted by a journal.}

Integrated nonlinear photonics has enabled compact and phase-coherent light sources for applications in precision spectroscopy, frequency metrology, optical frequency division, and microwave generation~\cite{stern2020direct, spencerOpticalfrequencySynthesizerUsing2018,newman2019architecture, sun2024integrated}. Many of these applications require coherent second harmonic generation (SHG) for their critical functions such as spectral extension for probing ultra-narrow atomic and molecular transitions and $f$-$2f$ self-referencing for frequency comb stabilization. Achieving stable and accurate SHG in a fully integrated platform is therefore of central importance in precision applications. Platforms that possess an intrinsic $\chi^{(2)}$ nonlinearity, such as lithium niobate~\cite{zhu2021integrated, boes2023lithium}, aluminum nitride~\cite{guo2016second}, and silicon carbide~\cite{lukin20204H}, are becoming increasingly mature and accessible. On the other hand, silicon nitride, though an amorphous thin film that does not possess a $\chi^{(2)}$ nonlinearity in bulk, has been shown to support SHG through the coherent photogalvanic (PG) effect. Here, a photo-induced DC electric field combines with the medium's $\chi^{(3)}$ nonlinearity to produce an effective $\chi^{(2)}$ response~\cite{porcel2017photo, billat2017large}. Silicon nitride is of particular interest due to its broad transparency window, high Kerr nonlinearity, ultra-low loss~\cite{blumenthal2018silicon}, CMOS compatibility~\cite{moss2013CMOS}, potential for hybrid laser integration~\cite{tran2022extending, heim2025hybrid} and established foundry fabrication on up to 300~mm wafers~\cite{ou2025300mm}. This maturity has enabled extensive research into nonlinear optical processes including Kerr frequency comb generation~\cite{gaeta2019photonic}, optical parametric oscillation~\cite{lu2026photonic}, second- and third-harmonic generation~\cite{levy2011harmonic}, and cascaded nonlinear optical processes~\cite{hu2022photo}. 

\begin{figure*}[htp]
    \centering
    \includegraphics[width=0.9\linewidth]{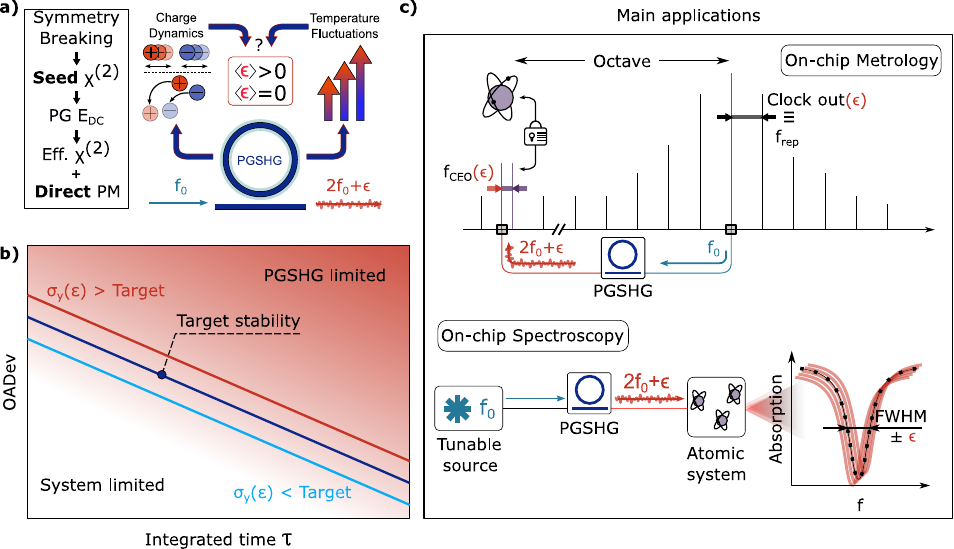}
    \caption{Highly stable photogalvanic second harmonic generation (PG-SHG): a) Schematic description of the PG-SHG physical mechanism based on direct phase-matching (PM) and potential additional noise $\epsilon$ introduced by the process, for example, due to charge dynamics associated with the DC electric field (E$_\text{DC}$) creation via the PG effect. b) Long-term stability requirement: the effect of $\epsilon$ is negligible only when it is masked by a target system whose instability (represented by the Overlapping Allan deviation - OADev) is sufficiently large. c) Any noise $\epsilon$ associated with PG-SHG can affect the stability of the optical clock output ($f_{rep}$ and carrier envelope offset frequency, $f_{CEO}$, is dependent on $\epsilon$), as well as the performance of high-end PG-SHG atomic spectroscopy. }
    \label{fig:concept}
\end{figure*}

PG-SHG can be achieved through two main phase-matching approaches: quasi-phase matching, where phase-mismatch between fundamental and second harmonic waves creates a spatial modulation of the DC electric field whose wavenumber automatically compensates the mismatch, and direct phase matching, where fundamental and second harmonic waves propagate in modes of different transverse order that have the same phase velocity. To achieve the efficiency needed for applications such as self-referencing of octave-spanning microcombs or frequency doubling of continuous wave lasers, resonant enhancement of SHG is essential. The above phase-matching strategies have both been implemented in microrings~\cite{lu2021efficient,nitiss2022optically}. The automatic quasi-phase matching approach~\cite{nitiss2022optically} has been particularly compelling with regards to its flexibility, with recent achievements including ultra-broadband SHG in reconfigurable coupled-ring structures~\cite{clementi2025ultrabroadband}, and its combination with self-injection locking to achieve low-phase-noise lasers near 780~nm and 532~nm~\cite{li2023high, wang2026integrated, yuan2025efficient}. Recently, however, work utilizing automatic quasi-phase matching has reported frequency shifts on the order of 10~Hz to 200~Hz and attributed them to the PG effect creating a spatio-temporal electric field pattern, with the shift value related to pump laser power and cavity detuning~\cite{zhou2025self}. Such offsets directly challenge the assumption that photogalvanic SHG preserves an exact $2{:}1$ frequency relationship and therefore raise a fundamental concern for its use in precision frequency conversion. It remains unknown whether these offsets are intrinsic to the photogalvanic mechanism, arise specifically from automatic quasi-phase matching, or can be avoided through direct phase matching. Consequently, the long-term frequency accuracy and metrological fidelity of resonantly enhanced PG-SHG remains unclear.

In this paper, we demonstrate metrologically compatible PG-SHG with direct phase matching in a silicon nitride microring~\cite{lu2021efficient} by verifying the (2{:}1) frequency-ratio fidelity. We measure the stability of the $2{:}1$ frequency ratio between the fundamental and doubled light and quantify the corresponding fractional frequency instability over extended timescales. We observe a residual instability at the $2\times10^{-15}$ level at 1~s integration time, averaging down to the $10^{-16}$ level at 1000~s, with the deviation from the exact doubling falling within the margin of the measurement uncertainty, and the induced nonlinaer response remaining stable for up to 5~hrs. Moreover, the offset observed due to the pump detuning is of the order of the measurement uncertainty, in contrast with the aforementioned approach using the automatic quasi phase-matching approach~\cite{zhou2025self}. These measurements establish an upper bound on the contribution of PG-SHG to frequency instability and systematic uncertainty, demonstrating that photogalvanic $\chi^{(2)}$ in silicon nitride can support phase-coherent and stable frequency doubling.  

\begin{figure*}[htp]
    \centering
    \includegraphics[width=0.95\linewidth]{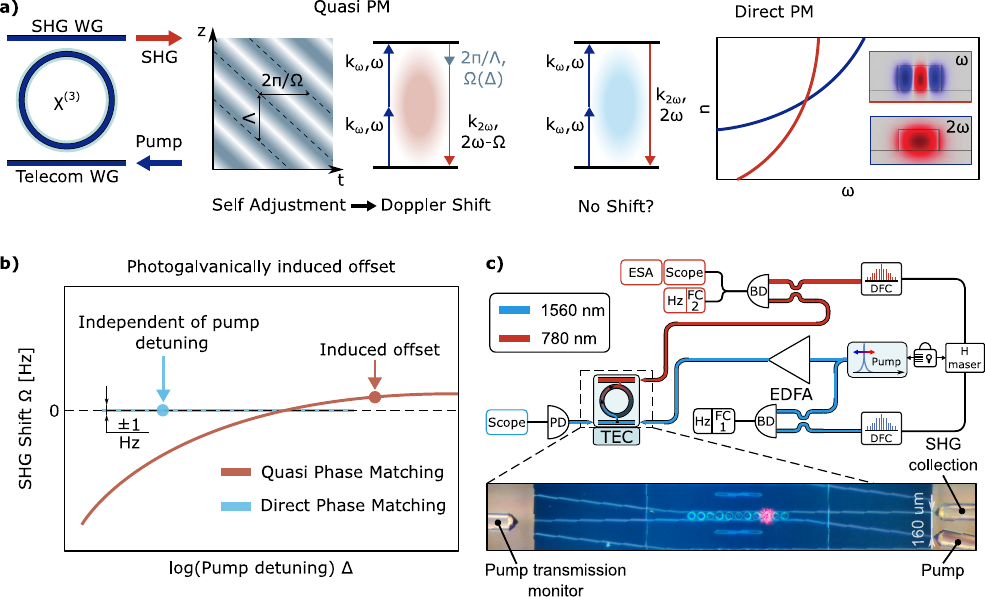}
    \caption{Photogalvanic phase matching and experimental setup. a) A silicon nitride microring resonator (MRR) in an add-drop configuration is used to generate photogalvanic second-harmonic generation (PG-SHG). In the quasi-phase-matching (QPM) approach, a dynamic spatio-temporal grating is formed, schematically adapted from Ref.~\cite{zhou2025self}. This grating introduces an additional degree of freedom in the energy- and momentum-conservation picture and can lead to pump-detuning-dependent offsets of the generated second-harmonic frequency. In the direct phase-matching approach, phase matching is achieved between two resonator modes of different order, avoiding the need for a self-organized QPM grating. Instead, we hypothesize that direct phase matching provides a detuning-independent frequency-doubling condition with no offset in the second harmonic frequency, and examine this hypothesis in this work. b) Schematic comparison between QPM-based PG-SHG, where the generated second-harmonic frequency can depend on pump detuning, and directly phase-matched PG-SHG, where the (2{:}1) frequency relation is expected to be preserved independently of pump detuning. This makes direct phase matching particularly attractive for compact optical atomic clocks and other precision frequency-conversion applications. c) Experimental setup. A pump laser (at 1560~nm) is amplified by an erbium-doped fiber amplifier (EDFA) and coupled into the MRR near resonance, resulting in PG-SHG (at 780 nm). Fiber-to-chip coupling for intracavity-power monitoring and SHG collection is performed using lensed fibers (spacing of 160 $\mu$m). The pump laser is stabilized to a maser-referenced difference-frequency comb (DFC). Balanced photodetectors are used to measure the pump and SHG beat notes against the DFC, and synchronized frequency counters (FCs) are used to extract the deviation ($\epsilon$) from exact doubling. An electrical spectrum analyzer (ESA) and an oscilloscope are used for electrical characterization of the beat signals. The inset shows a photograph of the chip during PG-SHG operation, with visible (780 nm) emission and lensed-fiber tails.}
    \label{fig:setup}
\end{figure*}

Before describing our experimental work in detail, we provide additional application context in Figure~\ref{fig:concept}. In SHG, the generated frequency can be evaluated as $2f_0 + \epsilon$, where $\epsilon$ captures all deviations from ideal doubling, including additional frequency noise and systematic offsets present in the experimentally obtained signal (see Fig.\ref{fig:concept} [a]). The discrepancy can arise from physical mechanisms, such as the dynamics of the charges which establish a DC field in the PG effect, as well as from experimental systematic contributions. We evaluate the stability and frequency fidelity of PG-SHG in a SiN microring resonator, focusing on the requirements imposed by future integrated systems for precision metrology and high-resolution spectroscopy. First, we assess the frequency accuracy of the process by measuring any systematic offset from exact doubling. Second, we evaluate the stability of the (2{:}1) frequency ratio over time, which determines whether the frequency uncertainty can average down during extended measurements. Third, we examine short-time phase noise and coherence preservation, which determine suitability for narrow-linewidth operation and high-resolution spectroscopy. 

The relevance of the above measurements on applications is illustrated in Fig.~\ref{fig:concept}(b)-(c). In optical atomic clocks (Fig.~\ref{fig:concept}(c), top), an optical frequency comb (OFC) is employed to transfer the relative stability from an optical frequency atomic reference to the OFC repetition rate (the microwave output)~\cite{fortier201920}. A critical element in this stability transfer, known as optical frequency division (OFD), is stabilization of the OFC carrier-envelope offset frequency ($f_\text{CEO}$) via self-referencing with SHG, where any additional noise introduced by the doubling procedure directly affects the clock output stability. In high-resolution spectroscopy (Fig.~\ref{fig:concept}(c),bottom), additional noise introduced in SHG can degrade coherence and compromise accurate linewidth measurements. We quantify the long-term noise behavior of any deviation in perfect SHG ($\epsilon$) using the Overlapping Allan Deviation (OADev) as shown in Fig.\ref{fig:concept}(b). By comparing the OADev of the PG-SHG process to the target stability, we can determine its impact. 

\begin{figure*}[htp]
    \centering
    \includegraphics[width=0.95\linewidth]{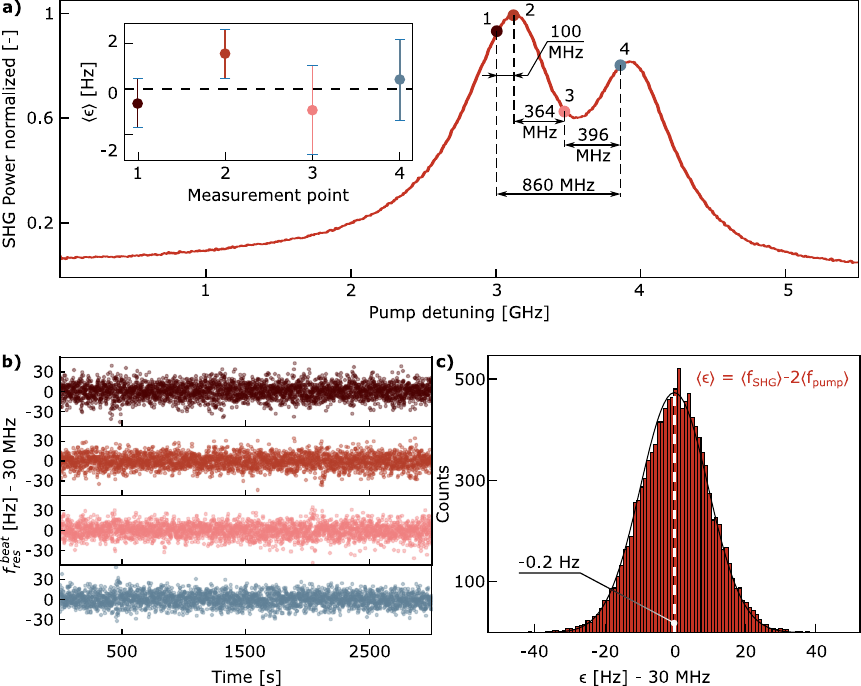}
    \caption{Photogalvanic SHG offset in direct phase-matching: a) SHG power dependence (normalized to the maximum SHG power) on the pump detuning. Four measurement points are identified by the solid circles, and denote detunings where $\langle\epsilon\rangle$ was measured (inset). b) Time traces of the measurements corresponding to the points shown in (a) from which the $\langle\epsilon\rangle$ values and accompanying error margins were obtained, where the error margins are the 95~\% confidence intervals of the Allan deviation. c) Histogram of the combined measurement (pure statistical type A evaluation) consisting of the 4 distinct points in (b), exhibiting a gaussian distribution with $\langle\epsilon\rangle = (-0.2~\pm~1.46)$~Hz, demonstrating the lack of correlation between the measurements.}
    \label{fig:epsilons}
\end{figure*}

Figure~\ref{fig:setup}(a)-(b) illustrates a potential role of the phase-matching approach in the above considerations. Quasi phase-matching produces a spatio-temporal grating that self-adjusts to the deviation between the pump and its second harmonic~\cite{zhou2025self}. This adjustment results in an additional spatial term $\Lambda$ and temporal term $\Omega$ in the energy-momentum conservation picture (Fig.~\ref{fig:setup}(a)), with $\Omega$ depending on the pump detuning with respect to its corresponding cavity resonance (Fig.~\ref{fig:setup}(b)). Direct phase-matching does not have an additional spatial term, and our work addresses whether an additional temporal term (and accompanying frequency shift) is observed. We do not measure any significant offset or dependence on the pump frequency (see Fig.\ref{fig:setup}[b]), suggesting that PG-SHG can be used for demanding metrology applications.

Our experimental setup is depicted in Fig.~\ref{fig:setup}[c]. The silicon nitride microring is designed to provide a direct phase matching between the pump resonance at frequency $f_0$ and its second harmonic at $2f_0$ with integrated coupling waveguides at both frequencies \cite{lu2021efficient}. A tunable laser amplified by means of an erbium-doped fiber amplifier (EDFA) is used as a pump. After second-harmonic generation is established, the pump laser is stabilized to a table-top difference-frequency comb (DFC) referenced to a hydrogen maser, suppressing pump-frequency excursions and enabling a cleaner measurement of residual deviations from exact frequency doubling. On-chip and off-chip coupling of light is performed by lensed fibers at 1550~nm and 780~nm, respectively. The large parameter space of the chip in terms of the number of available devices imposes constraints on efficient input and output coupling. Specifically, the distance of 160~$\mu$m between the 1560~nm lensed fiber and the 780~nm lensed fiber (see Fig.~\ref{fig:setup}[c]) forces, in our current realization, the part of the fibers (up to 5 mm) to be suspended in the air, which might contribute to the systematic measurement noise. The photonic chip is mounted on a thermo-electric cooler (TEC), which is used to control the temperature of the MRR. Finally, balanced detectors are used to monitor the beat notes of the pump and SHG with respect to the stabilized DFC, as well as for linewidth and phase noise measurements.

\begin{figure*}[htp]
    \centering
        \includegraphics[width=\linewidth]{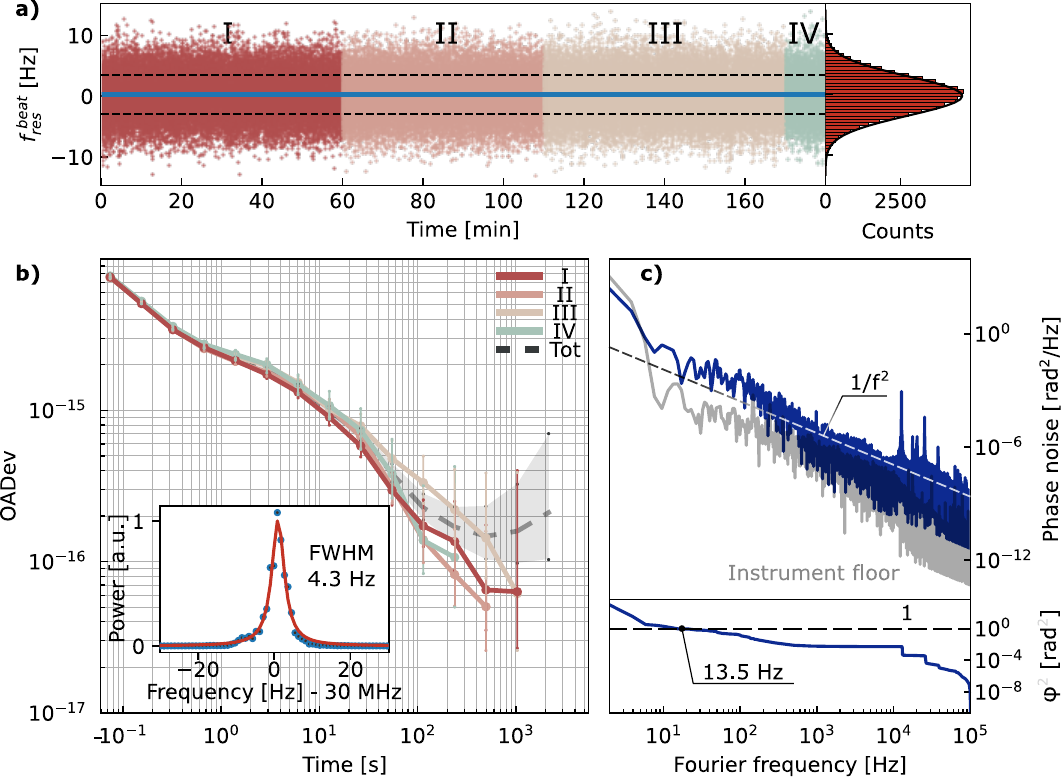}
    \caption{PG-SHG stability and accuracy measurement: (a) Time trace of the SHG beat frequency, exhibiting a standard deviation of 3.26 Hz. The four segments correspond to successive relocking events. (b) Overlapping Allan deviation (OADev) of the SHG beat frequency; each color corresponds to a segment in (a), while the grey dashed line shows the combined result. The vertical bars indicate the uncertainty of each individual run, while the gray shaded region indicates the uncertainty (99~\% confidence interval) of the combined run. The inset shows the beat note obtained by the ESA, which when fit to a Lorentzian, has a FWHM of (4.3~$\pm$~0.012)~Hz, where the one standard deviation uncertainty is obtained from the covariant matrix of the nonlinear fit. (c) Phase noise of the SHG beat (top, blue trace), together with the corresponding integrated linewidth (bottom), yielding a value of 13.5 Hz at 1 rad$^2$.}
    \label{fig:results}
\end{figure*}

Our experimental setup is depicted in Fig.~\ref{fig:setup}[c]. The silicon nitride microring is designed to provide a direct phase matching between the pump resonance at frequency $f_0$ and its second harmonic at $2f_0$ with integrated coupling waveguides at both frequencies \cite{lu2021efficient}. A tunable laser amplified by means of an erbium-doped fiber amplifier (EDFA) is used as a pump. After second-harmonic generation is established, the pump laser is stabilized to a table-top difference-frequency comb (DFC) referenced to a hydrogen maser, suppressing pump-frequency excursions and enabling a cleaner measurement of residual deviations from exact frequency doubling. On-chip and off-chip coupling of light is performed by lensed fibers at 1550~nm and 780~nm, respectively. The large parameter space of the chip in terms of the number of available devices imposes constraints on efficient input and output coupling. Specifically, the distance of 160~$\mu$m between the 1560~nm lensed fiber and the 780~nm lensed fiber (see Fig.~\ref{fig:setup}[c]) forces, in our current realization, the part of the fibers (up to 5 mm) to be suspended in the air, which might contribute to the systematic measurement noise. The photonic chip is mounted on a thermo-electric cooler (TEC), which is used to control the temperature of the MRR. Finally, balanced detectors are used to monitor the beat notes of the pump and SHG with respect to the stabilized DFC, as well as for linewidth and phase noise measurements.

To initiate PG-SHG, we set our pump resonance at 1556.9 nm (300 mW input power) with second harmonic output generated at 778.45 nm. Upon successful SHG, the microring emits bright red light (100 $\mu$W output power including coupling losses) as is shown in Fig.~\ref{fig:setup}[c]. Multiplication of the pump frequency by a factor of 2 due to PG-SHG implies (see supplementary materials) that the ratio between the two monitored beats (pump and SHG) is also equal to 2. The pump is monitored prior to entering the chip, while the SHG is measured at the output, such that the measurement isolates contributions arising from the nonlinear frequency conversion itself and from experimental systematics, both encompassed by the $\epsilon$ term. The DFC used for beat-note generation and the associated locking electronics are both referenced to the same hydrogen maser, suppressing reference-chain inconsistencies.

To identify the specific offset between the pump laser and the MRR resonance that gives the highest output SHG power and for the further measurement of the SHG frequency dependence on pump detuning, we scanned the pump laser and simultaneously monitored the SHG collection port (see Fig.~\ref{fig:epsilons}[a]). The double-peak shape of the SHG response strongly suggests that we are operating at perfectly phase-matched conditions \cite{roland2016phase}. First, we set the pump detuning a few MHz to the high-frequency side of the resonance peak, achieving highly stable SHG operation sustained for up to 5 hours. Second, to demonstrate that the direct phase-matching is largely independent of pump detuning, we investigated four distinct points across the SHG curve, as shown in Fig. \ref{fig:epsilons}(a). The overall pump detuning among all the measurement points reached 860 MHz, which is a significant portion of the whole direct phase-matched operation range as can be seen from the plot. At each point the pump and SHG residual beat notes were monitored simultaneously using synchronized frequency counters (FCs), and the resulting time traces are shown in Fig.~\ref{fig:epsilons}[b]. From the time traces obtained it was possible to compare the mean value of the SHG with the mean value of the pump, which is essentially the mean value of $\epsilon$. The values of $\langle \epsilon \rangle$ are shown in the inset of Fig.~\ref{fig:epsilons}[a] with their corresponding error margins, and are (-0.6~$\pm$~1.07)~Hz, (1.5~$\pm$~1.08)~Hz, (-0.92~$\pm$~1.93)~Hz, and (0.42~$\pm$~1.77)~Hz (see Table 1). The error margin is defined as the 95~\% confidence interval of the Allan Deviation (ADev), representing the statistical uncertainty of the frequency stability measurements. This measurement demonstrates that for direct phase-matching in PG-SHG, the pump detuning does not induce an offset in the second harmonic exceeding the measurement uncertainty—a clear contrast to the QPM case. The histogram of the combined measurement consisting of all 4 measurement points (see Fig.~\ref{fig:epsilons}[c]) shows an even smaller $\langle\epsilon\rangle$ of (-0.2~$\pm$~1.46)~Hz, emphasizing the lack of correlation between the different points.

\begin{table}[]
    \centering
    \begin{tabular}{c|c|c}
     Measurement & Absolute offset [Hz] & Uncertainty [Hz]\\
     I & -0.6 & $\pm$ 1.068\\ 
     II & 1.5 & $\pm$ 1.08\\
     III & -0.92& $\pm$ 1.93\\
     IV & 0.42& $\pm$ 1.77\\
\end{tabular}
    \caption{Measured frequency offset of the PG-SHG process as a function of cavity detuning.}
    \label{table1}
\end{table}

Further, we performed a long-term measurement of the SHG residual noise $\epsilon$ to understand its suitability for demanding metrological applications like optical atomic clocks. The different colors in Fig.~\ref{fig:results}[a] denote measurement segments of \{60, 40, 60, 20\} minutes, respectively. This segmentation arises from occasional loss of lock and subsequent relocking. A standard deviation of 3.26 Hz is obtained over a total measurement time of 3 hours, while the longest PG-SHG operation obtained could last up to 5 hours. The overlapping Allan deviation (OADev) obtained from the time traces characterizes the residual stability of three consecutive measurement runs and of their combined record. The uncertainty of each individual run is indicated by vertical bars, while the uncertainty of the combined record is shown by the pale gray shaded region, and corresponds to a 99~\% confidence interval. The mean value of the residual frequency trace yields an absolute deviation from exact frequency doubling of 0.5$~\mathrm{Hz}$. 
In an $f-2f$ clockwork, any residual deviation from exact doubling, $\epsilon(t)=f_{\mathrm{SHG}}(t)-2f_n(t)$, appears directly as an additive error on the measured carrier-envelope-offset beat. The mean value of $\epsilon$ therefore represents a possible systematic bias in $f_{\mathrm{CEO}}$, while its temporal fluctuations represent an added instability of the clockwork. In our measurement, the mean residual offset of $0.5~\mathrm{Hz}$ would correspond to a $0.5~\mathrm{Hz}$ bias of the $f$-to-$2f$ beat, or a fractional optical-frequency error of $1.3\times10^{-15}$ when referred to $780~\mathrm{nm}$. The measured instability of $\epsilon$, reaching $3.3\times10^{-15}$ at $1~\mathrm{s}$ and averaging down to $4.5\times10^{-16}$ at $1000~\mathrm{s}$, therefore places an upper bound on the instability that PG-SHG would contribute to an integrated $f-2f$ clockwork.

The obtained measurement, derived from the concatenation of all four segments, is compatible with the best known miniaturized optical atomic references\cite{klinger2025cs, kitching2024next, andeweg2026active} operating at $\approx10^{-14}$ relative frequency instability level at 1~s of the integration time. Furthermore, our result accommodates this benchmark with a sufficient margin, supporting the potential for even lower instability levels. The coherence transfer of the SHG process on shorter time scales has been further characterized by measuring the linewidth of the residual signal and its phase noise. The inset of Fig.~\ref{fig:results}[b] shows the beat linewidth obtained from the ESA (resolution bandwidth of 1~Hz), where a close Lorentz approximation results in a full-width at half maximum (FWHM) of (4.3~$\pm$~0.012)~Hz. This measurement is consistent with the obtained time trace standard value and indicates that the measurement system is dominated by white frequency noise, that is additionally backed up by a phase noise measurement with a clear $1/f^2$ trend. From the phase noise, we obtained the integrated linewidth at 1~rad$^2$ of 13.5 Hz, consistent with all the other measurements. Considering any additional unidentified systematic noise of the measurement related to the properties of the detectors and electronic amplifiers involved, such a result is readily compatible with the requirements of a $<1$~kHz-level spectroscopy \cite{legaie2018sub} and well beyond.

While the stability characteristics have been demonstrated to be compatible with the future integrated optical atomic clock architecture, we have also addressed the question of accuracy. To verify the absence of a constant frequency shift in case of a long-term measurement we have calculated the difference between the mean value of the SHG beat ($\bar{f}_{SHG})$ and doubled pump beat ($2\times \bar{f}_{pump})$, as we did previously for different pump detunings. For the 3 hour measurement of Fig.~\ref{fig:results}, we obtained an offset value of 0.63~Hz, which is statistically insignificant with respect to the 0.5 Hz measurement uncertainty extracted from the OADev and is in agreement with the values obtained before. It is nevertheless instructive to put the hard upper bound on the achievable stability transfer, which corresponds to $1.63\times10^{-15}$ for a 780 nm carrier (see Table 1 for each individual measurement). 

\begin{table}[]
    \centering
    \begin{tabular}{c|c|c}
     Measurement & Absolute offset & Relative offset \\
     I & 0.52 [Hz] & $1.36 \times 10^{-15}$ \\ 
     II & 0.32 [Hz] & $0.84 \times 10^{-15}$ \\
     III & 0.97 [Hz] & $2.53 \times 10^{-15}$ \\
     IV & 0.64 [Hz] & $1.66 \times 10^{-15}$ \\
     Total & 0.63 [Hz] & $1.63 \times 10^{-15}$ \\
\end{tabular}
    \caption{Measured frequency offset of the PG-SHG process relative to ideal 2:1 frequency doubling. Total stands for the combined measurement, denoted in gray in Fig.~\ref{fig:results}.}
    \label{table2}
\end{table}

It is also worth mentioning that along with Silicon Nitride other material platforms demonstrate exceptional performance in non-linear conversion. For example, periodically poled lithium niobate has emerged as a leading platform for second-harmonic generation~\cite{zhu2021integrated, boes2023lithium}. However, while wafer-level fabrication of lithium niobate has advanced rapidly, driven largely by industrial demand for optical modulation, wafer-level poling remains challenging, in part because wafer-thickness variations require adapted poling technologies~\cite{lithium_wafer_1, chen2024adapted} and because high-power operation can be limited by photorefractive effects~\cite{sayem2026high}. As a result, wafer-level periodic poling has not yet become a standard scalable foundry process, and multi-project-wafer (MPW) access to periodically poled lithium niobate remains limited. Other material platforms, including aluminum nitride~\cite{guo2016second} and silicon carbide~\cite{lukin20204H}, have also been explored, but their thin-film growth and fabrication processes remain less standardized and scalable than those of lithium niobate. Thus, even from the perspective of the fabrication maturity and availability of MPW runs, Silicon Nitride is the material of choice. 

Given the growing interest in PG-SHG as a route to integrated self-referencing~\cite{zhou2025self, clementi2025ultrabroadband, clementi2023chip, lu2021efficient} for octave-spanning microcombs that can interface with optical frequency standards near 780~nm~\cite{moilleAllopticalNoiseQuenching2025}, this work verifies that directly phase-matched PG-SHG in Silicon Nitride can preserve the $2{:}1$ frequency relation with the fidelity required for an optical clockwork. We show that the absolute deviation from ideal doubling remains at the $<1$~Hz level over the explored pump-detuning range, with no resolvable detuning-dependent offset within the 1~Hz level measurement uncertainty. Over a three-hour measurement, the residual fractional instability reaches $3.3\times10^{-15}$ at $1~\mathrm{s}$ and averages down to $4.5\times10^{-16}$ at $1000~\mathrm{s}$. We further find that the $2{:}1$ frequency ratio is maintained with a residual offset of ($0.63\pm0.5)~\mathrm{Hz}$, corresponding to a fractional optical-frequency error of $1.63\times10^{-15}$. These measurements establish directly phase-matched PG-SHG in SiN as a promising route for integrated self-referencing, compact optical clockworks, and frequency-calibrated precision spectroscopy. The measured low residual phase noise and integrated linewidth of $13.5~\mathrm{Hz}$ further demonstrate the suitability of PG-SHG for high-coherence clockwork and spectroscopy applications. Overall, these results establish an important metrological benchmark for miniaturized integrated optical atomic clocks and precision spectrometers, where $\chi^{(3)}$ and effective $\chi^{(2)}$ nonlinearities can coexist on a mature CMOS-compatible silicon nitride platform.\\

\noindent \textbf{Acknowledgments} The authors thank Jordan Stone and Gr\'egory Moille for helpful conversations. X.L. and K.S. acknowledge the NIST-on-a-chip program for funding support. NIST work was funded solely by the U.S. government. X.L. also acknowledges Maryland Industrial Partnerships for funding support. The fabrication was mainly performed at the NIST Center for Nanoscale Science and Technology, and in part at the Cornell NanoScale Facility, funded by the National Science
Foundation (NNCI-2025233).\\

\noindent \textbf{Data availability} The data that support the findings of this study are available from the corresponding author upon reasonable request.
\newpage

\bibliography{PG-OPO}

\end{document}


\section{Photo-induced Second Harmonic Generation Procedure}

After coupling the amplified pump laser to the MRR at the telecom wavelength around the MRR resonance, we initiate the pump laser sweep of approximately 20 GHz around the resonance and simultaneously monitor the output power at the SHG port. In order to initiate the PG-SHG process, it is necessary to adjust the polarization of the pump such that the basic $TE_{00}$ mode of the MRR is coupled. By gradually moving the pump wavelength while being continuously swept, we observe the characteristic SHG peak appearing at the SHG output on the scope. It takes a few (5~s to 10~s) in order for a peak to be formed for the first time after the idle operation and to reach a steady-state. Concurrently, we monitor the chip with a microscope and the initiation of PG-SHG is accompanied by emerging red light, pulsating according to the frequency of the pump sweep. After obtaining the characteristic ideal phase-matching SHG peak for the first time, the re-initiation process is instantaneous. It is then possible to identify the specific pump offset a few MHz aside from the SHG peak, and by red-shifting the pump to this specific point, to initiate a stable PG-SHG operation with a characteristic bright red light emitted from the chip. 

\section{Derivation of the beat relation between the fundamental and the second harmonic}

Here, we want to derive the ratio between the two frequency beats of the pump laser and the SHG light to isolate the residual noise components. We define the first beat frequency, $f_{b1}$, as the offset between the pump laser and the $N$-th harmonic of the repetition rate of the DFC around 1560 nm:
\begin{equation}
    f_{b1} = f_{\text{pump}} - N f_r \equiv \Delta
\end{equation}
The second beat frequency, $f_{b2}$, corresponds to the difference between the SHG light and the corresponding closest tooth of the DFC around 780 nm. Accounting for the PG-SHG induced frequency fluctuation $\epsilon$, the relationship is expressed as:
\begin{equation}
    f_{b2} = (2 + \epsilon) f_{\text{pump}} - 2N f_r = (2 + \epsilon)(\Delta + N f_r) - 2N f_r = 2\Delta + \epsilon \Delta + \epsilon N f_r
\end{equation}

From these expressions, we can determine the ratio $\alpha$:
\begin{equation}
    \alpha = \frac{f_{b2}}{f_{b1}} = 2 + \epsilon + \frac{\epsilon N f_r}{\Delta}
\end{equation}
Alternatively, we define the difference $D$ as:
\begin{equation}
    D = f_{b2} - 2 f_{b1} = \epsilon \Delta + \epsilon N f_r \approx \epsilon N f_r
\end{equation}
This difference $D$ yields a larger, more measurable quantity and serves as an excellent approximation for the frequency offset (in Hz) of the SHG from exact frequency doubling. 

To evaluate the stability of this system, we analyze the fluctuations $\delta D$. The total error in $D$ is given by the expansion:
\begin{equation}
    \delta D = \delta \epsilon \Delta + \epsilon \delta \Delta + \delta \epsilon N f_r + \epsilon N \delta f_r \approx \delta \epsilon N f_r
\end{equation}
In this expansion, the terms are of vastly different magnitudes: the first is $O(10^{-8})$, the second $O(10^{-18})$, the third $\sim 1$, and the final term $O(10^{-13})$. Consequently, the fluctuation is dominated by the $\delta \epsilon N f_r$ term. This behavior is mirrored in the fluctuations of the second beat:
\begin{equation}
    \delta f_{b2} = 2 \delta \Delta + \delta \epsilon \Delta + \epsilon \delta \Delta + \delta \epsilon N f_r + \epsilon N \delta f_r \approx \delta \epsilon N f_r
\end{equation}
This approximation holds under the condition $2 \delta \Delta \ll 1$, which is standard for systems under tight phase-locking. Nevertheless, it is technically superior to analyze $D$ directly by measuring $f_{b1}$ and $f_{b2}$ in correlation.

Finally, we relate these terms to the total noise of the SHG signal. The frequency of the SHG light is:
\begin{equation}
    f_{\text{SHG}} = (2 + \epsilon)(\Delta + N f_r)
\end{equation}
The resulting noise contribution is described by:
\begin{equation}
    \delta f_{\text{SHG}} = \delta f_{b2} + 2 N \delta f_r
\end{equation}
By utilizing this correlated measurement approach, we effectively eliminate the common-mode pump noise and isolate the residual noise $\epsilon$ inherent to the conversion process.